# SHARPENING SKILLS IN USING PRESENTATION TOOLS: STUDENTS' EXPERIENCES

Maria Angeles Dano-Hinosolango

Department of Technology Communication Management
Mindanao University of Science and Technology
Cagayan de Oro City, Philippines
angiedano28@gmail.com

*ABSTRACT*

*Making use of Information and Communication Technology resources is deemed necessary for students. With this, they need to demonstrate their competence in using various technologies such as Prezi and PowerPoint to communicate effectively and efficiently. Hence, their experiences in using these presentation tools is important to assist them in their needs. In this study, all the fourth year Bachelor of Science in Technology Communication Management students who have used the PowerPoint and Prezi were the respondents. Survey instruments were given to determine the experiences of students of these technologies based on their familiarity, skills, and effectiveness in delivering the reports and presentations. It was found out that students were generally good in using these tools. On the other hand, following the basic rules to use these presentation tools effectively has to be reinforced in the classroom to enhance and enrich their learning in sharpening their skills on presentation tools.*

## 1. BACKGROUND AND RATIONALE

Making use of information and communication technology (ICT) is important in presenting reports and presentations for the students to enrich their learning. Hence, they need to demonstrate their competence in using various technologies such as Prezi and PowerPoint (PPT) to communicate effectively and efficiently to their audience. With this, they can integrate multimedia platforms as tools in delivering a particular topic. Thus, their experiences in using these presentation tools is important to assist them in their needs that will consequently aid in enriching and sharpening their skills on giving reports and presentations.

Microsoft's PowerPoint presentation tool allows the students to integrate multimedia features. It is not only limited to text, graphics, animations and transitions but also equip in embedding audio, videos and music for the audience to understand better the topic presented. It was shown that if PPT is used appropriately it becomes a powerful tool that will encourage students' learning [1]. In like manner, Prezi is a web-based presentation tool that also allows the users to integrate multimedia features. Its difference with PPT is it allows the presenter to magnify the text, to make the elements upside down, and the like. It was also shared that the "theory behind Prezi is that our ideas are not linear, but rather bundles of interconnected concepts that are better captured as a whole with many parts. With this, it allows the presenter to illustrate the relationship of concepts to one another" [2].

On the other hand, there is a challenge of being competent and effective in using these tools. For PPT, it might be too wordy or having a lot of images in a slide that could distract the audience. In Prezi, a lot of animations are going on that might lead the audience at lost. In this connection, students' familiarity of the presentation tools, skills in using PPT and Prezi and effectiveness in their delivery shall be considered as variables in this study. As students, it is important for them to be equipped with various ICT presentation tools to communicate effectively and efficiently their reports and presentations. It is in this light that the objective of this study is to determine the experiences of the students in PPT and Prezi in the classroom to equip them with their needs that will aid in enriching the course content in sharpening their skills in using presentation tools.

## 2. STUDENTS' EXPERIENCES IN USING THE PRESENTATION TOOLS

The students' experiences in using the presentation tools can serve as a guide in assisting them in their needs on how to deliver effectively and efficiently their reports and presentations in the classroom. Henceforth, the following salient points were considered in this study.

### 2.1. Theoretical Background and Brief Literature

This study is anchored on the Cognitive Theory of Multimedia Learning and Constructivist Learning Theory.

Cognitive Theory of Multimedia Learning states that learning "is an active process of filtering, selecting, organizing, and integrating information" as cited by Doyle (2011) from the work of (Mayer, 2009) [3]. It presents the "idea that the brain does not interpret a multimedia presentation of words, pictures, and auditory information in a mutually exclusive fashion; rather, these elements are selected and organized dynamically to produce logical mental constructs. Furthermore, it underscores the importance of learning (based upon the testing of content and demonstrating the successful transfer of knowledge) when new information is integrated with prior knowledge." In relation to the use of presentation tools like PPT and Prezi, this theory affirms that the multimedia skills are important among students in order to enhance their presentational skills and to enrich the learning of their audience. One also made mentioned that the "knowledge conveyed to the listeners increases when the presenter's style is dynamic and engaging" [4].

In like manner, Constructivism Learning Theory presents that human beings produce or construct meaning, understanding and knowledge of the world from their own experiences. Constructivist epistemology assumes that students produce or construct their own meaning or knowledge based on their interactions with their environment. A related path to constructivism comes from Gestalt theories of perception (Kohler, 1925) that focus on the ideas of closure, organization, and continuity (Bower & Hilgard, 1981). Like Vico, Gestalt psychologists suggested that people do not interpret pieces of information separately and that cognition imposes organization on the world. With this, students are encouraged to create their own PPT and Prezi integrating their multimedia skills based on their understanding of the assigned report and presentation. Thus, students are able to construct new meaning by choosing the flow of their presentation, lay outing through a graphic organizer, embedding audio, videos and music, etc. to enrich their presentation.

Moreover, it was also mentioned that "as a presenter in front of an audience, it is key that the information has meaning, but it also organized in a way that it can be easily understood and absorbed by the audience. It is crucial to have their best interests at hand when it comes to information design. Also important is the presenter having control over the information and the tools being used to present that information so as to create the best experience possible for the audience" [5]. Thus, in this study students were asked on their experiences regarding familiarity, skills and effectiveness in using PPT and Prezi. For familiarity, students identified how familiar and often they used the tools in the classroom; for skills, they determined how often they used the features as well as the difficulties encountered; and for effectiveness, it measured their content and delivery in making and using these presentation tools.

## 2.2. Methodology

The study was done at Mindanao University of Science and Technology in the second semester of academic year 2013-2014. It made use of the nonprobability sampling procedure specifically, the purposive sampling. This was conducted to all the fourth year students who were taking up Bachelor of Science in Technology Communication Management (BSTCM) who used both the PPT and Prezi in their reports and presentations. During the Multimedia and Professional Presentation classes, these tools were introduced and taught to them as part of the course content. This provided them the experiences to use these presentation tools.

The survey instruments were given to the students at the end of the semester. The researcher-made questionnaires was given separately, one for the PPT and one for Prezi. This was given separately for the purpose that they were not to compare these presentation tools since the focus of the study was to look at their experiences for each of the tool that would aid in enriching the course content. It also made use of focused-group discussions (FGD) to substantiate the results of the study.

To interpret the results, descriptive research design was used. The data were analyzed by using the descriptive statistics such as percentage, mean and standard deviation to describe the experiences of the BSTCM students in using PPT and Prezi based on their familiarity, skills and effectiveness in giving reports and presentations.

## 2.3. Highlights of Findings and Discussion

This study attempted to determine and analyze the experiences of the students in using PPT and Prezi in the classroom based on their familiarity, skills and effectiveness in using these presentation tools.

Table 1 shows the level of familiarity of using PPT and Prezi in the classroom. In presenting reports and presentations, majority of the respondents most often used PPT. On the other hand, majority of the students only sometimes used Prezi.

During the FGD, students shared that they used PPT most of the time because of its accessibility. They shared that they only sometimes used Prezi because of limited internet connection. This also shows some hard truths that some students in some areas of Northern Mindanao have limited access to internet connection. On the other hand, since students were familiar with PPT, they already explored most of its features and did not dwell with the ready-made templates. For Prezi, there were moments that they used ready-made designs. This implies that students became familiar as they explored and used these tools in their reports and presentations.

After the discussion in the Multimedia and Professional courses about the basic rules in presenting a good and effective PPT and Prezi, they most often applied the rules in PPT because they mentioned that they also learned along the way from the comments of their peers and teachers. On the other hand, in Prezi students shared that there were moments that they still needed to apply the basic rules because sometimes they were overwhelmed with its features and focused on it instead of the content. This is similar to what has been cited by White (2011) from the work of Adria (2009) that whizzing could be distracting that might lose the essence of the message being shared to the audience [5].

Table 1. Percentage Distribution on the Familiarity of PowerPoint and Prezi

| | Familiarity | | |
|---|---|---|---|
| | | **Percentage** | |
| **Indicators** | | **PPT** | **Prezi** |
| How often do you use these tools in presenting reports? | • sometimes used | 13.33 | 80.00 |
| | • most often used | 66.67 | 20.00 |
| | • at all times used | 20.00 | 0.00 |
| How familiar are you with the features of PPT and Prezi? | • just used the ready-made templates | 0.00 | 33.33 |
| | • most often used the various features | 43.33 | 36.67 |
| | • maximized the use of various features | 56.67 | 30.00 |
| How often do you apply the basic rules in using these tools? | • sometimes applied | 20.00 | 50.00 |
| | • most of the time applied | 46.67 | 40.00 |
| | • at all times applied | 33.33 | 10.00 |

Table 2.1 presents the overall skills of the students in using the features of PPT and Prezi particularly on the effects in animations and transitions, inserting shapes and objects, graphs, pictures, music and videos, and other features like hyperlinking. In like manner, Table 2.2 identifies to what degree they have demonstrated the skills in using the features of these presentation tools.

In Table 2.1, it shows the assessment of students regarding their own skills to what extent they have used the features in these presentation tools. One also shared that "self-assessment means more than students grading their own work; it means involving them in the processes of determining what is good in any given situation" [6]. Hence, self-assessment serves as a stepping stone to enrich one's skills.

Table 2.1 Skills in Using the Following Features of PowerPoint and Prezi

| Features | | PowerPoint | | | Prezi | | |
|---|---|---|---|---|---|---|---|
| | | **Mean** | **SD** | **Desc** | **Mean** | **SD** | **Desc** |
| Effects | Animations/Transitions | 4.77 | 0.43 | Very Good | 4.13 | 0.97 | Good |
| Inserting | Shapes/Objects | 4.53 | 0.63 | Very Good | 3.87 | 1.40 | Good |
| | Graphs | 4.20 | 0.81 | Good | 3.37 | 1.40 | Fair |
| | Pictures | 4.80 | 0.48 | Very Good | 4.20 | 0.92 | Good |
| | Music | 4.00 | 1.17 | Good | 2.97 | 1.27 | Fair |
| | Videos | 3.67 | 1.35 | Good | 2.93 | 1.39 | Fair |
| Others | Hyperlinking | 2.93 | 1.39 | Fair | 2.30 | 1.29 | Needs Improvement |

* Legend: Desc = Description

It reveals that they were generally skilled in PPT. They only needed to improve their skills on hyperlinking. In Prezi, they were generally good but fair in some areas like inserting graphs, music and videos. Students also shared that in Prezi they found it quite challenging to hyperlink because some sites would not open without internet connection. With this, it limits their opportunity to use the hyperlink feature.

Table 2.2 presents the skills demonstrated in using the features in these presentations tools. This particularly focused on choosing templates, font style and size, blending of colors, animations and transitions, inserting graphs and the like.

For the design, it shows that majority of the students already designed their own templates in PPT. In Prezi, they used ready-made templates and design their own frames. This was a good indicator since students became more creative in designing their own slides and frames that would fit in their respective presentations.

In choosing the font style and size, majority of the students always modified them in making PPT, and most of them also did the same in Prezi. This is very important for the whole audience to clearly read the text even at the back of the classroom. They also shared that they tried their best not to condense all the texts in the slides or frames to avoid too wordy and quite heavy presentations. With regard to font choice, one found out in his study that "Gill Sans, a popular sans serif font, was rated highly on each of the four variables, making it a safe choice for PowerPoint slides. These four variables considered were comfortable-to-read, professional, interesting, and attractive" [7].

Taking into consideration blending of the right colors, majority of the students explored first the right contrast of hues. Students shared that they needed to choose the right primary colors that would blend to the texts and background. This helped them to make their presentation better and clearer to the audience. However, there were also a number of students who have difficulty in blending the right colors. They shared that it really took them time to choose the right color of text and background. There were also times that they took it for granted. Hence, students can be taught of the skill in matching colors for texts and backgrounds.

For animations and transitions, most of the respondents already obtained the skill in choosing these particular feature in PPT. On the other hand, in Prezi most of them used only some animations and transitions. It is very important that students should use the right animations and transitions that would not be distracting to the audience. This feature should enhance the presentation not to distract or simply impress the audience. With regard to Prezi, the "pathing" skills should be planned well in using graphic organizers so that the audience would not get lost and see the interconnection of ideas. It was shared that "Prezi has faced some less than gleaming reviews for its unfamiliar interface and dizzying zooming capabilities" [5]. Thus, the skill on using animations and transitions shall be taken into consideration.

In using shapes and objects, majority of the students already applied the skill of inserting shapes and objects in their slides. They shared that they usually used this for headings in each slide. It guided their audience which part they were in their presentations. In Prezi, majority of them have also used this feature. Some of the students inserted these in frames to enhance the presentation. They also shared that they both used this in PPT and Prezi especially if they used images in order to add texts. It would not be clear if they overlap texts with pictures or images.

With regard to inserting graphs, most of them have inserted graphs both in PPT and Prezi. They shared that this was necessary in their Thesis proposal and final defense. "Graphs greatly increases the clarity of presentation and makes it easier for a reader to understand the data being used and to draw clear and correct inferences" [8]. On the other hand, a number of students also shared that they experienced difficulty because Excel asked them regarding figures. They shared that this is their waterloo. It took them time to get the right formula in order to come up with the right distribution in their graphs. With this, students can be taught on the basic formula to get the right figures in the graphs.

Table 2.2 Percentage Distribution on the Students' Skills in Using the Following Features in PowerPoint and Prezi

| Features | | Percentage | |
|---|---|---|---|
| | Indicators | PPT | Prezi |
| Design | • no idea how to look for design/templates | 0.00 | 0.00 |
| | • use ready-made designs/templates | 6.67 | 36.67 |
| | • look for more designs in the Internet | 6.67 | 16.67 |
| | • modify ready-made designs/templates | 23.33 | 23.33 |
| | • design my own templates | 63.33 | 23.33 |
| Font Style and Size | • no idea how to choose the font style/size | 0.00 | 0.00 |
| | • use ready-made font style/size | 6.67 | 16.67 |
| | • seldom modify the font style/size | 3.33 | 10.00 |
| | • often modify the font style/size | 20.00 | 30.00 |
| | • always modify the font style/size | 70.00 | 43.33 |
| Blending of Colors | • no idea how to blend the colors well | 3.33 | 3.33 |
| | • have some knowledge | 3.33 | 6.67 |
| | • have some difficulties | 6.67 | 13.33 |
| | • explore appropriate colors | 56.67 | 63.33 |
| | • have the skills on blending the right colors | 30.00 | 13.33 |
| Animations and Transitions | • no idea how to use them | 0.00 | 3.33 |
| | • have some knowledge | 10.00 | 20.00 |
| | • have some difficulties | 0.00 | 6.67 |
| | • have used some animations and transitions | 43.33 | 46.67 |
| | • have the skills in choosing the right animations and transitions | 46.67 | 23.33 |
| Shapes/ Objects | • no idea how to insert shapes/objects | 0.00 | 0.00 |
| | • have some knowledge | 10.00 | 16.67 |
| | • have some difficulties | 3.33 | 10.00 |
| | • have inserted shapes/objects | 36.67 | 60.00 |
| | • have the skills on inserting the shapes/objects | 50.00 | 13.33 |
| Graphs | • no idea how to insert graphs | 0.00 | 6.67 |
| | • have some knowledge | 20.00 | 23.33 |
| | • have some difficulties | 10.00 | 13.33 |
| | • have inserted graphs | 40.00 | 43.33 |
| | • have the skills on inserting graphs | 30.00 | 13.33 |
| Pictures/ Images | • no idea how to insert pictures | 0.00 | 0.00 |
| | • have some knowledge | 10.00 | 10.00 |
| | • have some difficulties | 3.33 | 16.67 |
| | • have inserted pictures/images | 26.67 | 33.33 |
| | • have the skills on inserting pictures/images | 60.00 | 40.00 |
| Hyperlinking | • no idea how to hyperlink | 13.33 | 26.67 |
| | • have some knowledge | 23.33 | 16.67 |
| | • have some difficulties | 10.00 | 13.33 |
| | • have tried including hyperlinks | 36.67 | 40.00 |
| | • have the skills on including hyperlinks | 16.67 | 3.33 |

For inserting pictures and images, majority of the students applied this in their slides. In Prezi, most of them also used this skill. Students shared that they have imported pictures, images, emoticons, and the like from their desktop or from Google. Pictures and images are important to enrich the presentation especially to show and illustrate something to the audience. Including pictures in presentations is a simple and powerful way of expanding one's expressive potential as a speaker. Moreover, they mentioned that "pictures communicate at levels beyond the descriptive possibilities of words and bathe the brain in much desired visual stimulation. At the same time, not all pictures are created equally. Choosing the right images, and using them in the right ways, can greatly impact your effectiveness" [9].

Lastly, in hyperlinking most of them have tried already using this feature both in PPT and Prezi. They liked hyperlinking because it is already embedded in the slides or frames, and they did not need to open videos, sites, etc. in their documents that could interrupt the flow of their presentations. On the other hand, there are limitations of this feature. Its disadvantage is it does not open all the hyperlinks. For instance, there is a hyperlink that needs internet connection and/or it does not support the files that are hyperlinked.

In general, students have applied the skills in using the features in PPT and Prezi. However, there are skills that need to be enhanced and enriched among TCM students.

Table 3 presents their skills on the effectiveness of their outline, content, delivery and performance in using PPT and Prezi. Students checked all the skills that they applied and experienced in making reports and presentations.

Table 3. Percentage Distribution on Using the PowerPoint and Prezi Effectively

| Features | | | |
|---|---|---|---|
| | Indicators | PPT | Prezi |
| Outlining | • never outline my content | 3.33 | 6.67 |
| | • rarely outline | 3.33 | 20.00 |
| | • sometimes outline | 23.33 | 13.33 |
| | • often outline | 36.67 | 33.33 |
| | • always outline | 33.33 | 26.67 |
| Content | • know the 7 by 7 rule | 66.67 | 30.00 |
| | • simply copy and paste the text | 20.00 | 26.67 |
| | • use bullets that serve as a guide in the presentation | 70.00 | 66.67 |
| | • rephrase/paraphrase/summarize significant points | 60.00 | 56.67 |
| | • only highlight significant points | 76.67 | 56.67 |
| Delivery | • simply read the text or content | 6.67 | 10.00 |
| | • read and explain the content | 60.00 | 66.67 |
| | • confident in delivering reports/presentations | 43.33 | 36.67 |
| | • comfortable presenting reports using PPT/Prezi | 73.33 | 40.00 |
| | • still use index cards or cue notes in reporting | 33.33 | 36.67 |
| Performance | • appreciated by my peers | 60.00 | 53.33 |
| | • appreciated by my teachers | 56.67 | 46.67 |
| | • earn good grades | 66.67 | 63.33 |
| | • do not earn good grades | 6.67 | 6.67 |
| | • not appreciated by my classmates nor teachers | 3.33 | 10.00 |

In outlining, most of the respondents often outlined their report. This is a good indicator because it guided their audience on the flow of their presentation. However, this has also to be reinforced in the classroom because there were still a number who did not practice this skill. One shared that outlining the activities is a helpful technique that allows students to be quickly informed of what is about to happen in class for that particular period [10].

For the content, in PPT the strong points were following the "seven by seven rule", using bullets that would eventually lead to highlighting only significant points. This shows a positive result because it shows consistency of their responses. If they have followed the "seven by seven rule" which means as much as possible only having seven words and seven lines in a slide, they would eventually use bullets and only highlight important points in the presentation. In Prezi, majority of the students applied bullets, rephrased/ paraphrased/summarized significant points and highlighted important points. This is also a good indicator because students practiced the skill of not to condense all the words in just one slide or frame. On the other hand, there were also some students who simply copied and pasted their report from various sources. They shared that it would be easier for them to do this when they ran out of time. Thus, reminding the students to be prepared ahead of time will help them a lot to come up with a better presentation in class.

With regard to delivery, it showed that majority of the students did not only read the content but also explained it at the same time. In using PPT and Prezi, this is a good indicator because it is one of the main purposes of presentation. Presenters should show and tell what is the topic all about. They also shared that they were comfortable using the PPT because it is accessible and easy to use. For Prezi, they became more confident because it was something new to them, and they liked the whizzing and zooming features. However, they also shared some challenges in using these presentation tools. For instance, they needed internet connection to import pictures and gif images online. They also needed the Wi-Fi and/or internet connection to make Prezi. Moreover, even if they were taught how to use and make these presentation tools, they could not do hands-on activities because of poor internet connection, and the Multimedia Lab has to be upgraded with these tools. With this, students dwelt on checking the tutorials available in YouTube for using these tools and did the hands-on at home or internet cafes. These were the challenges and difficulty they encountered in making their reports and presentations.

Lastly, in rating their performance, it was rewarding for the students to be appreciated by their peers and teachers. They were well-appreciated if they made very good and effective PPT and Prezi. In addition, majority of them also earned good grades for their presentations. They shared that their peers and instructors observed that they improved in making their presentations. Some of the feedback were following the "seven by seven rule", including images instead of words, and choosing the right designs or templates. A study was conducted regarding an assessment of student preferences for PPT. It was revealed that "studies have demonstrated that students prefer PowerPoint and respond favorably to classes when it is used. Students preferred the use of key phrase outlines, pictures and graphs, slides to be built line by line, sounds from popular media or that support the pictures or graphics on the slide, color backgrounds, and to have the lights dimmed. It is recommended that professors pay attention to the physical aspects of PowerPoint slides and handouts to further enhance students' educational experience" [11]. On other hand, some of the students also shared that they were not appreciated or not received good grades because they overdo their transitions and animations, too wordy with their slides or frames that could hardly be read, and the like. With this, there are still skills that need to be improved and practiced more in their Multimedia and Professional Presentation courses.

## 3. CONCLUSIONS

Part of the vision and mission of the University is to form students equipped with the technological knowledge and skills. With the ASEAN 2015 and global trends in education, making use of ICT resources is deemed necessary for the digital natives to maximize the use and educational benefits of technology in the classroom like the presentation tools particularly PPT and Prezi. Hence, their experiences in using these presentation tools is important to assist them in their needs of using these technologies.

In this study, it was found out that students were generally good in using these tools. On the other hand, hands-on experience with an instructional guide to be used inside and outside the classroom is encouraged. Moreover, the basic rules to use these presentation tools effectively has to be reinforced in the classroom. This provides an opportunity to enrich their learning and be equipped with the use of these technologies.

With this, it is encouraged that the administrators shall provide more training on ICT resources to its faculty. If budget warrants, providing Wi-Fi and/or faster internet connection would be very beneficial. For the TCM department, enriching the course content and making instructional materials would be beneficial to students for them to be more guided on the effective and efficient use of PPT and Prezi. For the students, it is encouraged to make use of these presentation tools to effectively and efficiently deliver their reports and presentations. In addition, providing peer assessments in their work would also be beneficial in order to enrich their multimedia and professional presentations skills.

## ACKNOWLEDGEMENTS

I would like to extend my deepest gratitude to the BSTCM students who participated in this study, my department chair Ms. Angeli Pizarro-Monsanto for her unceasing support, our Research Head Dr. Amparo Vedua-Dinagsao for her valuable comments and suggestions, and my husband Glenn and daughter Zyrhene for their love and understanding.

**Author**

Dr. Maria Angeles D. Hinosolango graduated Cum Laude in March 2001 with the degree Bachelor of Secondary Education major in English. She finished her Master of Arts in Education major in Communication Arts (English) and Doctor of Philosophy in Education at Xavier University – Ateneo de Cagayan.

Before taking up her doctorate degree, she was granted a Fulbright scholarship as a Foreign Language Teaching Assistant (FLTA) at Northern Illinois University on August 2008-May 2009.

She taught in Xavier University High School for a decade. At present, she is teaching in the Department of Technology Communication Management at Mindanao University of Science and Technology.

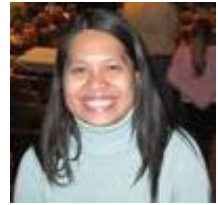